\documentclass{mn2e}
\usepackage{lmodern}

\usepackage[T1]{fontenc}
\usepackage[utf8]{inputenc}
\usepackage{color}
\usepackage{amsmath}
\usepackage{amssymb}
\usepackage{graphicx}

\usepackage{hyperref}
\hypersetup{hidelinks}

\title[Boosting MC sampling with a non-Gaussian fit]{Boosting MonteCarlo sampling with a non-Gaussian fit}

\author[L. Amendola and A. G\'omez-Valent]{
Luca Amendola\thanks{l.amendola@thphys.uni-heidelberg.de}
and Adrià Gómez-Valent\thanks{gomez-valent@thphys.uni-heidelberg.de}
\\
Institut f\"{u}r Theoretische Physik, Ruprecht-Karls-Universit\"{a}t Heidelberg, Philosophenweg 16, D-69120 Heidelberg, Germany
}

\begin{document}
\label{firstpage}
\pagerange{\pageref{firstpage}--\pageref{lastpage}}
\maketitle

\begin{abstract}
We propose a new method, called MonteCarlo Posterior Fit, to boost the MonteCarlo sampling of likelihood (posterior) functions. The idea is to approximate the posterior function by an analytical multidimensional non-Gaussian fit. The many free parameters of this fit can be obtained by a smaller sampling than is needed to derive the full numerical posterior. In the examples that we consider, based on supernovae and cosmic microwave background data, we find that one needs an order of magnitude smaller sampling than in the standard algorithms to achieve comparable precision. This method can be applied to a variety of situations and is expected to significantly improve the performance of the MonteCarlo routines in all the cases in which sampling is very time-consuming. Finally, it can also be applied to Fisher matrix forecasts, and can help solve various limitations of the standard approach. 
\end{abstract}


\maketitle

\section{introduction}

A large fraction of today's  cosmology consists in finding the probability distribution of parameters (called posterior in Bayesian language) given a set of observational data. Since both the cosmological models  and the data points are an ever growing set, the development of fast algorithms to sample the posterior is a clearly perceived need. A typical MonteCarlo (MC) sampling for Cosmic Microwave Background (CMB) data might require hundreds of thousands of points to cover the dozens of parameters (nuisance plus cosmological) that adequately model the theory and the experiment systematics. This, in turn, might require days of computation, which should be repeated for every competing model. If the comparison with theory involves very time-consuming operations, like for instance $N$-body simulations, this task could quickly become unfeasible.
                                                                                                                                                  
In this paper we propose an alternative approach that, under certain circumstances, may reduce the computational times by one or more orders of magnitude. The idea is extremely simple: instead of sampling the posterior with as many points as possible to reconstruct it faithfully, we propose to sample the posterior a limited number of times and use these points to fit it with a flexible analytical distribution based on a higher-order expansion over a multivariate Gaussian. The sampling of the fitted distribution can be done with the same MonteCarlo algorithms usually employed, but it can be evaluated much more rapidly than the original (exact) posterior in the cases of interest, since it is fully analytical. The great advantage of our method is that, as we show in this paper, the overall computational time spent in the total sampling process can be one order of magnitude smaller than with the traditional method, with a minimal loss of precision in estimating the confidence regions. We call our method MonteCarlo Posterior Fit. For other interesting approaches, based on Gaussian Processes, see (Leclercq 2018, McClintock \& Rozo 2019, Pellejero-Ibañez et al. 2019) and references therein.

Although we perform in this work the sampling with the Metropolis-Hastings algorithm (Metropolis et al. 1953, Hastings
1970) and a minimal variant of it as explained in Sec. \ref{sec:Marginalization}, our method can be also used with other (sometimes more efficient) sampling techniques in order to boost even more the calculations. Some possibilities are:  Nested Sampling (Skilling 2004), Affine Invariant MonteCarlo Markov Chain (Goodman \& Weare 2010), Hamiltonian MonteCarlo (Duane et al. 1987), Density estimation likelihood-free inference (Fan et al. 2012; Papamakarios \& Murray 2016), or Approximate Bayesian Computation (Ishida et al. 2015; Akeret et al. 2015) among others.

Our method is very general and by no means limited to cosmological data, although here we only discuss illustrative examples based on cosmology, in particular Supernovae Ia (SnIa) and CMB. The code to implement our method, \texttt{MCPostFit}, is publicly available\footnote{\textcolor{blue}{\url{https://github.com/adriagova/MCPostFit}}}.

The method we propose finds another application as a generalization of the Fisher matrix (FM) approach. The Fisher matrix is a simple way to approximate the posterior, mostly employed for forecasting the performance of future observations given an expected theoretical model (see for example many applications in Amendola et al. 2018). In this case,  the best fit parameters are known in advance, and coincide with the assumed model. There are however three main problems with the standard FM approach. First, the FM, being given by the second derivative around the best fit point, is insensitive to the features away from the peak and is therefore by construction  a good approximation only in its vicinity (the non-Gaussian generalization explored by Sellentin 2014 and Sellentin, Quartin \& Amendola 2015 improve upon this but remains a local approximation). Secondly, the FM assumes that the best fit is a well-behaved peak, with zero first derivatives  and no flat directions. Thirdly, often the derivatives have to be taken numerically by taking small differences, not analytically, and small numerical errors might get amplified and distort the results. Our method, as we will show, solves all three problems. In a sense, our method fills the gap between the Fisher matrix method, which is based on a few points around the peak, and the full MonteCarlo sampling based on hundreds of thousands of points across the parameter space.

The crucial ingredient of our method is that the non-Gaussian fit to the posterior is characterized by a (typically) large  number of parameters  (we call this a "second-parametrization", to distinguish the posterior fit parameters from the theoretical ones) which, however, appear linearly in the fit function. This makes it possible to obtain a  simple {\it analytical} solution to the fitting problem, without any need of an expensive search through the very high-dimensional space of second-parameters. As we will show, the number of second-parameters might well be of the order of thousands for a typical CMB application, yet without any significant overhead in computational time.


\section{A MonteCarlo Gaussian fit to the posterior}\label{sec:G}

The central idea of our method is to  set up  an   approximation of the posterior obtained through a fit with a limited number of sampling points. In this section the fit will be taken to be  Gaussian, while in the next section we generalize to a non-Gaussian function. Notice that we never require either the data or the posterior to be Gaussian.

Let us denote with $L(\theta)$ a posterior function (for simplicity and with a slight abuse of language we sometimes refer to this function as the likelihood) that depends on a  vector of theoretical parameters $\theta$ living in a $N$-dimensional space. Let us then generate
a number $N_{r}$ of random $\mathbf{\theta}_{i}$ vectors, and evaluate $L_{i}\equiv L(\mathbf{\theta}_{i})$, through a MonteCarlo Markov Chain (MCMC) algorithm.
We denote with $\hat L$ the peak  of the posterior obtained for the best fit $\hat\theta$. We assume for now that the peak has been accurately determined, but later on we remove this assumption.
We can  then find a  Gaussian fit to the posterior by minimizing the quantity
\begin{equation}\label{eq:chi2}
\chi_{M}^{2}=\sum_{i=1}^{N_r}(\log L_{i}-\log L_{M}(\theta_{i}))^{2}
\end{equation}
 with respect to  the symmetric and constant\footnote{Any dependence of $M_{\alpha\beta}$ on $\theta$ would break the Gaussian nature of $L_M$, of course. We study such dependence in Sec. \ref{sec:NG}, where we introduce perturbatively some non-Gaussian corrections.} matrix $M_{\alpha\beta}$, where (sum over repeated indexes)
\begin{equation}\label{eq:fitGaussian}
L_{M}(\theta)=\hat{L}\exp\left[-\frac{1}{2}(\theta-\hat{\theta})_{\alpha}M_{\alpha\beta}(\theta-\hat{\theta})_{\beta}\right]\,.
\end{equation}
We have then
\begin{equation}
\log L_{M}=\log\hat{L}-\frac{1}{2}(\theta-\hat{\theta})_{\alpha}M_{\alpha\beta}(\theta-\hat{\theta})_{\beta}\,.
\end{equation}
We need then to solve the equation
\begin{equation}
\frac{\partial\chi^{2}_M}{\partial M_{\mu\nu}}=-2\sum_{i=1}^{N_r}(\log L_{i}-\log L_{M,i})\frac{\partial\log L_{M,i}}{\partial M_{\mu\nu}}=0\,,
\end{equation}
where $L_{M,i}=L_{M}(\theta_{i})$ and where $M_{\alpha\beta}$ are our $N(N+1)/2$ independent second-parameters. 
We find that  $\chi^{2}_M$ is extremized for
\begin{equation}
0=\sum_{i=1}^{N_r}(\log L_{i}-\log L_{M,i})(\theta_{i}-\hat{\theta})_{\mu}(\theta_{i}-\hat{\theta})_{\nu}\,,
\end{equation}
or, equivalently, 
\begin{equation}
\sum_{i=1}^{N_r}D_{\mu\nu,i}\log L_{M,i}=\sum_{i=1}^{N_r}D_{\mu\nu,i}\log L_{i}\,,
\end{equation}
where $D_{\mu\nu,i}\equiv(\theta_{i}-\hat{\theta})_{\mu}(\theta_{i}-\hat{\theta})_{\nu}$.
That is 
\begin{equation}
M_{\alpha\beta}\sum_{i=1}^{N_r}D_{\alpha\beta,i}D_{\mu\nu,i}=-2\sum_{i=1}^{N_r}D_{\mu\nu,i}\log\left(\frac{L_{i}}{\hat{L}}\right)\,.
\end{equation}
Let us now collect the pairs $\alpha,\beta$ into a single vector
of dimension $N(N+1)/2$.
E.g., if $\alpha,\beta$ run over $1,3$, we define $a=\{11,12,13,22,23,33\}$.
Then we can rewrite $M_{\alpha\beta}$ as $M_{a}$ (where for
the off-diagonal components $M_{a}=2M_{\alpha\beta}$) and $D_{\mu\nu}$
as $D_{a}$. Therefore,
\begin{equation}
M_{\alpha\beta}\sum_{i=1}^{N_r}D_{\alpha\beta,i}D_{\mu\nu,i}=M_{a}\sum_{i=1}^{N_r}D_{a,i}D_{b,i}\equiv M_{a}\Delta_{ab}\,.
\end{equation}
Then the solution is
\begin{equation}\label{eq:resultM}
M_{a}=2\Delta_{ab}^{-1}\sum_{i=1}^{N_r}D_{b,i}\log\left(\frac{\hat{L}}{L_{i}}\right)\,.
\end{equation}
Notice that we do not need to normalize the likelihood.   The number $N_r$ of points must be larger than $N(N+1)/2$ (otherwise the problem is underdetermined and the matrix $M$ is singular) and one should reach a $N_r$ large enough that  the solution $M_a$ converges, i.e., it does no longer significantly change by further increase of $N_r$ (we discuss this point further below). This number of posterior evaluations, $N_r$, should be compared to the number of evaluations of a typical MCMC algorithm. In all the applications below, we find that $N_r$ can be one order of magnitude, or more, smaller than a standard MCMC run, without appreciable loss of precision.


However, the posterior peak $\hat L$ and the best fit point $\hat\theta$ obtained with a limited number of samplings might not be very accurate. A better estimation can be obtained e.g. through the standard Newton-Raphson method (cf. Appendix \ref{sec:AppendixA}), but here instead we proceed in a much more efficient way, by generalizing our method to include the best fit parameters in our second-parametrization, still assuming that a Gaussian approximation to the posterior is sufficient.

 We consider again \eqref{eq:fitGaussian}, but now we minimize \eqref{eq:chi2} not only  w.r.t. the matrix elements $M_{\alpha\beta}$, but also  w.r.t. the components of the vector $\hat{\theta}$. Doing so however mixes the unknowns in a  non-linear way and the system cannot be longer solved analytically. We  perform therefore a transformation that makes the minimization procedure analytical again. Let us consider the following change of variables: $\hat{\theta}\to \tilde{\theta}+\Delta\theta$, with $\Delta\theta$ being the shift from the best fit vector in our Markov chain, $\tilde{\theta}$, to the optimal location of the fitting Gaussian's peak, $\hat{\theta}$. They will not coincide in general. We obtain,

\begin{multline}\label{eq:costFunction2}
\chi^2_M= \sum_{i=1}^{N_r}\bigg[\log\left(\frac{L_i}{\hat{L}}\right)+\frac{1}{2}M_{\alpha\beta}\Delta\theta_{\alpha}\Delta\theta_{\beta}-M_{\alpha\beta}\Delta\theta_{\alpha}(\theta_{i}-\tilde{\theta})_\beta\\
       +\frac{1}{2}M_{\alpha\beta}(\theta_{i}-\tilde{\theta})_\alpha(\theta_{i}-\tilde{\theta})_{\beta}\bigg]^2\,,
\end{multline}
which can be rewritten as follows,
\begin{multline}\label{eq:chi2MII}
\chi^2_M= \sum_{i=1}^{N_r}\bigg[\log\left(\frac{L_i}{\tilde{L}}\right)+\frac{1}{2}P_{\alpha}(\theta_{i}-\tilde{\theta})_\alpha\\
+\frac{1}{2}M_{\alpha\beta}(\theta_{i}-\tilde{\theta})_\alpha(\theta_{i}-\tilde{\theta})_\beta\bigg]^2\,,
\end{multline}
with 
\begin{equation}\label{eq:relations}
\tilde{L}=\hat{L}\exp\left[-\frac{1}{2}M_{\alpha\beta}\Delta\theta_\alpha\Delta\theta_\beta\right] \quad {\rm and}\quad P_\alpha = -2M_{\alpha\beta}\Delta\theta_{\alpha}\,.
\end{equation}
In practice, we have rewritten \eqref{eq:fitGaussian} just as
\begin{multline}\label{eq:fitGaussianImproved}
L_M(\theta)=\bar{L}\exp\bigg[-\frac{1}{2}(C+P_\alpha(\theta-\tilde{\theta})_\alpha\\
+M_{\alpha\beta}(\theta-\tilde{\theta})_\alpha(\theta-\tilde{\theta})_\beta)\bigg]\,,   
\end{multline}
where $\bar{L}e^{-C/2}=\tilde{L}$. With this simple change of variables we gain a lot, since now we can minimize $\chi^2_M$  w.r.t. $C$ ($\bar{L}$ can be fixed e.g. to $L(\tilde{\theta})$, which is known), the elements of the $N$-dimensional vector $P$ and, of course, the elements of the matrix $M$, solving equations that are fully linear in the second-parameters. The $N+1$ degrees of freedom that were contained in $\hat{L}$ and $\hat{\theta}$ are now contained in $C$ and $P$. The addition of the independent and linear terms in \eqref{eq:fitGaussian} allows us to correct the position and height of the peak without the need of any iterative routine, just applying the same formalism presented before. Minimizing \eqref{eq:chi2MII} w.r.t. $C$, $P$, and $M$ we obtain,
\begin{align}
\sum_{i=1}^{N_r}(C+P_\alpha D_{\alpha,i}+M_{\alpha\beta}D_{\alpha\beta,i}) & =2\sum_{i=1}^{N_r}\log\left(\frac{\bar{L}}{L_{i}}\right) \label{eq:systemGauss1}\\
\sum_{i=1}^{N_r}(C+P_\alpha D_{\alpha,i}+M_{\alpha\beta}D_{\alpha\beta,i})D_{\mu,i} & =2\sum_{i=1}^{N_r}\log\left(\frac{\bar{L}}{L_{i}}\right)D_{\mu,i}\\
\sum_{i=1}^{N_r}(C+P_\alpha D_{\alpha,i}+M_{\alpha\beta}D_{\alpha\beta,i})D_{\mu\nu,i} & =2\sum_{i=1}^{N_r}\log\left(\frac{\bar{L}}{L_{i}}\right)D_{\mu\nu,i}\,, \label{eq:systemGauss3}
\end{align}
where we have used the definitions
\begin{align}
D_\alpha & = (\theta-\tilde{\theta})_\alpha  \label{eq:def1}\\
D_{\alpha\beta} & =(\theta-\tilde{\theta})_{\alpha}(\theta-\tilde{\theta})_{\beta}\,. \label{eq:def2}
\end{align}
Then we can write the three equations \eqref{eq:systemGauss1}-\eqref{eq:systemGauss3} as
\begin{equation}
A\mathcal{M}=B\,,
\end{equation}
with 
\begin{align}
\mathcal{M} & =\left(\begin{array}{c}
C\\
P_{\alpha}\\
M_{a}
\end{array}\right)
\end{align}

\begin{equation}
 A= \sum_{i=1}^{N_r} \begin{pmatrix}
1  & D_{\alpha,i}  & D_{a,i} \\
D_{\beta,i} & D_{\alpha,i} D_{\beta,i} & D_{a,i} D_{\beta,i} \\
D_{b,i} & D_{\alpha,i}D_{b,i} & D_{a,i}D_{b,i} 
\end{pmatrix}  
\end{equation}

\begin{align}
B & =2\sum_{i=1}^{N_r}\log\left(\frac{\bar{L}}{L_{i}}\right)\left(\begin{array}{c}
1\\
D_{\beta,i}\\
D_{b,i}
\end{array}\right)
\end{align}
The solution is then
\begin{equation}\label{eq:M}
\mathcal{M}=A^{-1}B\,.
\end{equation}
One can use \eqref{eq:relations} and \eqref{eq:M} to compute the corrected position of the peak,
\begin{equation}
\hat{\theta}_\alpha=\tilde{\theta}_\alpha-\frac{1}{2}M^{-1}_{\alpha\beta} P_\beta\,.
\end{equation}
Even adding $\hat{L}$ and $\hat{\theta}$ among the second-parameters, the fit remains of course Gaussian. In the next section we will extend the method to take into account the possible non-Gaussianity of the posterior, by adding higher-order terms to the second-parametrization.

\begin{figure}
\includegraphics[width=8cm]{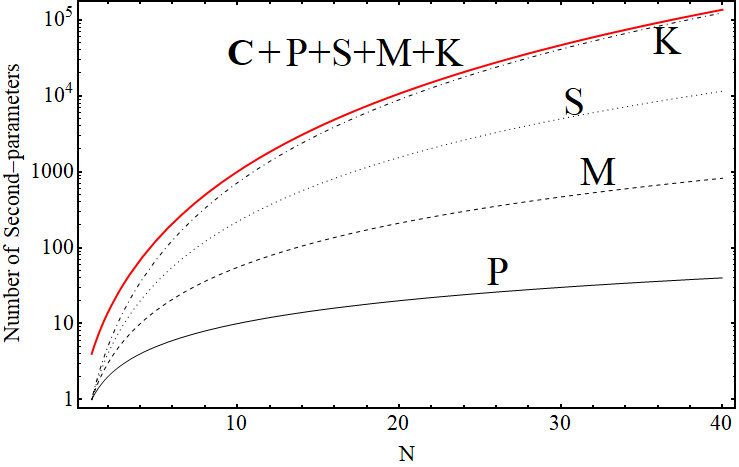}
\caption{\label{fig:dimA} Number of second parameters contained in each term of Eq. \eqref{eq:calM} as a function of $N$, when we include all the possible elements in each of these terms. We also show in red the total number of second-parameters contained in the vector $\mathcal{M}$, which is equal to the dimension of the square matrix $A$ \eqref{eq:A_matrix}.}
\end{figure}


\section{Non-Gaussian posterior fit}\label{sec:NG}

The method can be directly generalized to improve upon the quadratic approximation. In fact, the only crucial requirement is that the second-parameterization remains linear in the second-parameters. Clearly, the more second-parameters we introduce, the more accurate the fit becomes, provided there are more sampling points than second-parameters (but of course still less than is needed for the traditional MC sampling, otherwise the advantages of the method vanish).

We could then adopt the following higher-order non-Gaussian expression
\begin{align}\label{eq:full-fit}
L_{M}=\bar{L}\exp\bigg[-\frac{1}{2}(&C+P_\alpha D_\alpha \\
&+M_{\alpha\beta}D_{\alpha\beta}+S_{\alpha\beta\gamma}D_{\alpha\beta\gamma}+K_{\alpha\beta\gamma\delta}D_{\alpha\beta\gamma\delta})\bigg]\,, \nonumber
\end{align}
where the first two terms, already introduced in the previous section, help to improve the best fit, while the last two terms, $S$ (for skewness) and $K$ (for kurtosis), model possible deviations from  Gaussianity. Beside \eqref{eq:def1}-\eqref{eq:def2}, we also defined 
\begin{align}
D_{\alpha\beta\gamma} & =(\theta-\tilde{\theta})_{\alpha}(\theta-\tilde{\theta})_{\beta}(\theta-\tilde{\theta})_{\gamma}\\
D_{\alpha\beta\gamma\delta} & =(\theta-\tilde{\theta})_{\alpha}(\theta-\tilde{\theta})_{\beta}(\theta-\tilde{\theta})_{\gamma}(\theta-\tilde{\theta})_{\delta}
\end{align}
In other words, we replaced the quadratic Gaussian exponent with a multivariate polynomial of order four; clearly, this can be extended to arbitrarily higher order, paying the price of increasing complexity. In practice, we find that the third or fourth order expansion is sufficient in all the applications we considered. This includes the fitting of typical banana-shape posteriors and other highly-non Gaussian shapes (cf. Secs. \ref{sec:SNIa} and \ref{sec:CMB}, and Appendix \ref{sec:AppendixExamples}). It is important to remark though that Eq. \eqref{eq:full-fit} is  only  able  to  fit posteriors to  within  some  limits,  for  instance with  just  one or  two  peaks. Dealing with more general multimodal posteriors would obviously require the use of higher-order terms in the expansion. Eq. \eqref{eq:full-fit} can also have problems  describing very strong non-Gaussianities or small structures, but higher-order corrections are not needed in the vast majority of current cosmological studies.

\begin{figure*}
\includegraphics[width=15cm]{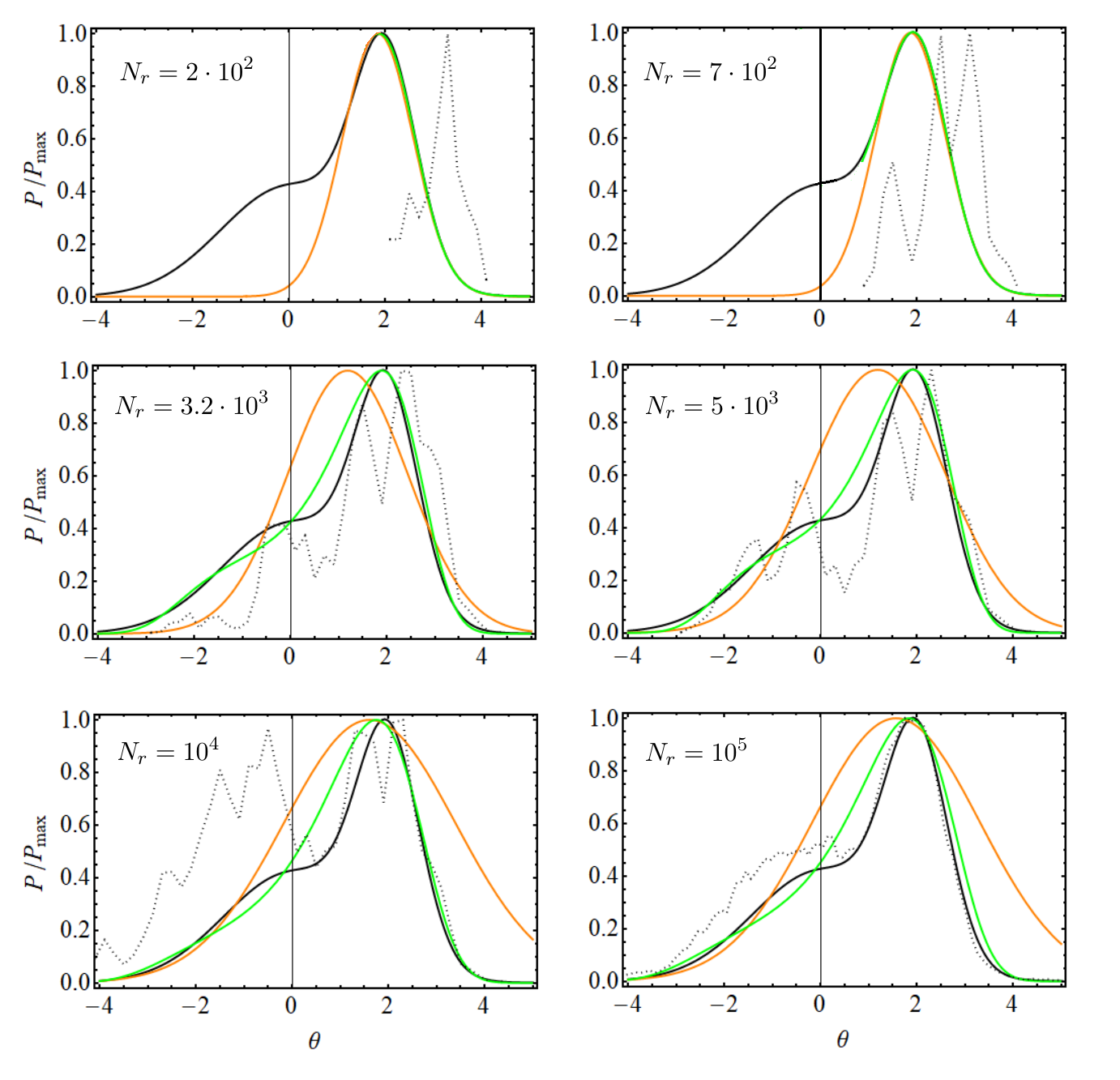}
\caption{\label{fig:fig3}
Example of 1D application of the MonteCarlo Posterior Fit. Black solid lines: exact distribution, sum of two Gaussians $\sim 0.5\,e^{-0.25x^2}+e^{-1.18(x-2)^2}$; Black dotted lines: histograms built from $N_r$ random points generated in the MonteCarlo, for 6 different values of $N_r$. For the proposal distribution to perform the jumps of the Metropolis-Hastings algorithm (Metropolis et al. 1953, Hastings 1970) we use a Gaussian with standard deviation $\sigma=0.1$; Orange line: MCFM Gaussian fit \eqref{eq:fitGaussianImproved}; Green line: MCFM non-Gaussian fit \eqref{eq:full-fit}. We have included in both cases the correction of the position of the peak, as described in Sec. \ref{sec:G}. See also the related comments in the main text of Sec. \ref{sec:NG}. }
\end{figure*}

The same procedure of minimization of Eq. (\ref{eq:chi2}) with respect to $C,P,M,S,K$ gives now 
the five equations
\begin{multline}
\sum_{i=1}^{N_r}(C+P_\alpha D_{\alpha,i}+M_{\alpha\beta}D_{\alpha\beta,i}\\ +S_{\alpha\beta\gamma}D_{\alpha\beta\gamma,i}+K_{\alpha\beta\gamma\delta}D_{\alpha\beta\gamma\delta,i})  = 2\sum_{i=1}^{N_r}\log\left(\frac{\bar{L}}{L_{i}}\right) 
\end{multline}
\begin{multline}
\sum_{i=1}^{N_r}(C+P_\alpha D_{\alpha,i}+M_{\alpha\beta}D_{\alpha\beta,i} +S_{\alpha\beta\gamma}D_{\alpha\beta\gamma,i} \\+K_{\alpha\beta\gamma\delta}D_{\alpha\beta\gamma\delta,i})D_{\mu,i} =  2\sum_{i=1}^{N_r}\log\left(\frac{\bar{L}}{L_{i}}\right)D_{\mu,i}
\end{multline}
\begin{multline}
\sum_{i=1}^{N_r}(C+P_\alpha D_{\alpha,i}+M_{\alpha\beta}D_{\alpha\beta,i}+S_{\alpha\beta\gamma}D_{\alpha\beta\gamma,i}\\ +K_{\alpha\beta\gamma\delta}D_{\alpha\beta\gamma\delta,i})D_{\mu\nu,i} = 2\sum_{i=1}^{N_r}\log\left(\frac{\bar{L}}{L_{i}}\right)D_{\mu\nu,i}
\end{multline}
\begin{multline}
\sum_{i=1}^{N_r}(C +P_\alpha D_{\alpha,i}+M_{\alpha\beta}D_{\alpha\beta,i}+S_{\alpha\beta\gamma}D_{\alpha\beta\gamma,i}\\ +K_{\alpha\beta\gamma\delta}D_{\alpha\beta\gamma\delta,i})D_{\mu\nu\sigma,i}  = 2\sum_{i=1}^{N_r}\log\left(\frac{\bar{L}}{L_{i}}\right)D_{\mu\nu\sigma,i}
\end{multline}
\begin{multline}
\sum_{i=1}^{N_r}(C+P_\alpha D_{\alpha,i}+M_{\alpha\beta}D_{\alpha\beta,i}+S_{\alpha\beta\gamma}D_{\alpha\beta\gamma,i}\\ +K_{\alpha\beta\gamma\delta}D_{\alpha\beta\gamma\delta,i})D_{\mu\nu\sigma\tau,i} = 2\sum_{i=1}^{N_r}\log\left(\frac{\bar{L}}{L_{i}}\right)D_{\mu\nu\sigma\tau,i}
\end{multline}
Extending the approach of the previous section, we organize now the indexes  as $\alpha\beta\to a=N(\alpha-1)+\beta$,
$\alpha\beta\gamma\to\bar{a}=N^{2}(\alpha-1)+N(\beta-1)+\gamma$,
$\alpha\beta\gamma\delta\to\tilde{a}=N^{3}(\alpha-1)+N^{2}(\beta-1)+N(\gamma-1)+\delta$, so we can write the five equations again as
\begin{equation}
A\mathcal{M}=B\,,
\end{equation}
where 
\begin{align}\label{eq:calM}
\mathcal{M} & =\left(\begin{array}{c}
C\\
P_{\alpha}\\
M_{a}\\
S_{\bar{a}}\\
K_{\tilde{a}}
\end{array}\right)
\end{align}

\begin{equation}\label{eq:A_matrix}
 A= \sum_{i=1}^{N_r} \begin{pmatrix}
1  & D_{\alpha,i}  & D_{a,i} & D_{\bar{a},i}  & D_{\tilde{a},i}\\
D_{\beta,i} & D_{\alpha,i} D_{\beta,i} & D_{a,i} D_{\beta,i} & D_{\bar{a},i} D_{\beta,i} & D_{\tilde{a},i} D_{\beta,i}\\
D_{b,i} & D_{\alpha,i}D_{b,i} & D_{a,i}D_{b,i} & D_{\bar{a},i}D_{b,i} & D_{\tilde{a},i}D_{b,i}\\
D_{\bar{b},i}  & D_{\alpha,i}D_{\bar{b},i} & D_{a,i}D_{\bar{b},i} & D_{\bar{a},i}D_{\bar{b},i} & D_{\tilde{a},i}D_{\bar{b},i}\\
D_{\tilde{b},i} & D_{\alpha,i}D_{\tilde{b},i} & D_{a,i}D_{\tilde{b},i} & D_{\bar{a},i}D_{\tilde{b},i} & D_{\tilde{a},i}D_{\tilde{b},i}
\end{pmatrix}  
\end{equation}

\begin{align}
B & =2\sum_{i=1}^{N_r}\log\left(\frac{\bar{L}}{L_{i}}\right)\left(\begin{array}{c}
1\\
D_{\beta,i}\\
D_{b,i}\\
D_{\bar{b},i}\\
D_{\tilde{b},i}
\end{array}\right)
\end{align}
(each entry in $\mathcal{M}$ and $B$ is a vector, each entry in $A$ is a block matrix).
The solution takes then the same form as in the Gaussian case:
\begin{equation}
\mathcal{M}=A^{-1}B\,.\label{eq:solng}
\end{equation}
The index organization is such that one considers only one permutation
of indexes (e.g. only $M_{12}$, and not $M_{21}$) and every entry
in $M,S,K$  acquires a factor equal to the number of
distinct permutations for that entry's indexes (e.g. in $112$, the
permutation of the 1s are not distinct). For instance, a combination
like $12$ takes a factor of $2!$ and therefore $2M_{12}\to M_{a}$;
a combination like $123$ a factor of $3!=6$ (i.e. $6S_{123}\to S_{\bar{a}}$);
a combination like $1122$, a factor of $4!/2!/2!=6$ (i.e. $6K_{1122}\to K_{\tilde{a}}$).
The ordering of the reduced indexes is arbitrary, but clearly once established
must be respected in every matrix. There are $(N+k-1)!/k!/(N-1)!$
elements, with $k=1,2,3,4$ for elements of type $\alpha,a,\bar{a},\tilde{a}$,
respectively.

The inversion of the matrix $A$ could become prohibitive for a large number of first-parameters, although there are nowadays very efficient algorithms to solve linear algebraic equations that avoid direct inversion. For instance, in a typical CMB analysis one could have around 30 free parameters (nuisance plus cosmological). The full $A$ matrix would then have dimensions of $46376\times 46376$, mostly due to the $K$ term (see Fig. \ref{fig:dimA}). However, there is no really need to include the same number of first-parameters in $P,M,S$ and $K$. For example, one could include all of them in $P,M$ and $S$ but a smaller subset (for instance only the cosmological ones) in  $K$, thereby drastically reducing the complexity.  We will experiment with these possibilities later on, in Sec. \ref{sec:CMB}.

In Fig. \ref{fig:fig3} we present  a simple artificial 1D application, using an underlying distribution which is highly non-Gaussian. It clearly shows the good ability of the method to reconstruct the posterior even when a small amount of MonteCarlo sampling points is employed in the fit, especially when non-Gaussian corrections are also considered in the analysis \eqref{eq:full-fit}. For such a small amount of sampling points, our method provides much better results than the standard MonteCarlo. Notice that the convergence is very fast. Already with only $\sim 3\%$ of the total number of points that are needed to get a good estimate of the posterior with the usual MonteCarlo approach we are able to obtain results that basically match with those obtained from the full MonteCarlo sample. In Sections \ref{sec:SNIa} and \ref{sec:CMB} we tackle more realistic cases. 

\section{Marginalization}\label{sec:Marginalization}

Once we have the full solution Eq. \eqref{eq:full-fit} we still need to marginalize over the nuisance parameters and over various subsets of parameters in order to produce 1D or 2D confidence regions. This can be obtained by sampling \eqref{eq:full-fit} with a standard Metropolis-Hastings algorithm (Metropolis et al. 1953, Hastings 1970), which however now is extremely fast since the posterior (\ref{eq:full-fit}) is analytical, which is not usually the case for the original distribution. Moreover, notice that, using the notation introduced earlier, Eq. \eqref{eq:full-fit} can be rewritten in the more compact form,
\begin{equation}
L_M=\bar{L}\exp\left[-\frac{1}{2}(C+P_\alpha D_\alpha+M_a D_a+S_{\bar{a}} D_{\bar{a}}+K_{\tilde{a}} D_{\tilde{a}})\right]\,,
\end{equation}
which can be evaluated much faster than \eqref{eq:full-fit}. We denote this sampling step as ``marginalization MC'' in order to clearly distinguish it from the first MC, which is carried out using the exact likelihood. It is important to keep in mind, though, that in general the fitted distribution $L_M$ will only describe accurately the true underlying distribution $L$ in the regions of parameter space explored in the first MC. Far away from these regions the reconstructed distribution can even diverge and the differences with respect to $L$ can be huge. In fact, there is no guarantee that the exponent of the likelihood is positive definite; in this case, there will always be regions far from the peak in which $L_M$ diverges. Fortunately, this problem can be controlled, minimizing its impact on the marginalization MCMC. One can start by restricting the generation of points in the marginalization MC inside the $N$-dimensional box set by the lowest and largest values of the parameters in each dimension. There can be still regions inside this box with a relatively low density of points in which the reconstructed distribution  begins to diverge. We can then evaluate the maximum value of $L_M$ around $\tilde{\theta}$, $L_M(\hat{\theta})$, and restrict the marginalization MC to those regions in which\footnote{In the Gaussian case we can easily compute $L(\hat{\theta})$ making use of \eqref{eq:relations}. In the non-Gaussian case there is no analytical expression for it, but we can estimate $L(\hat{\theta})$ by generating a list of random points around $\tilde{\theta}$ (say $\sim 10^4$ points) and then selecting the resulting maximum value. This computation only takes few minutes, so it is completely feasible.} $L_M<L_M(\hat{\theta})$.  If we take a point that does not fulfill this condition then we can force the MC to go back to the good region, by selecting randomly one of the $N_r$ sampling points obtained in the first MC. To control the jumping process we can employ a multivariate Gaussian, with the covariance matrix computed directly from the $N_r$ sampling points. It is also useful to shorten the steps by dividing the standard deviations by a certain factor, e.g. 10. In this way we can remain more time inside the region where the fitted distribution is well-behaved and improve the efficiency of the method. We have checked that this approach works very well (cf. Sec. \ref{sec:CMB}). The marginalization MC can be stopped as usual, once the well-known Gelman-Rubin convergence diagnostic (Gelman \& Rubin 1992) is below the desired threshold, typically $\hat{R}-1<10^{-3},10^{-2}$.  From the generated MonteCarlo Markov chain we can build then the corresponding histograms and compute the confidence regions.

\begin{figure*}
\includegraphics[width=15cm]{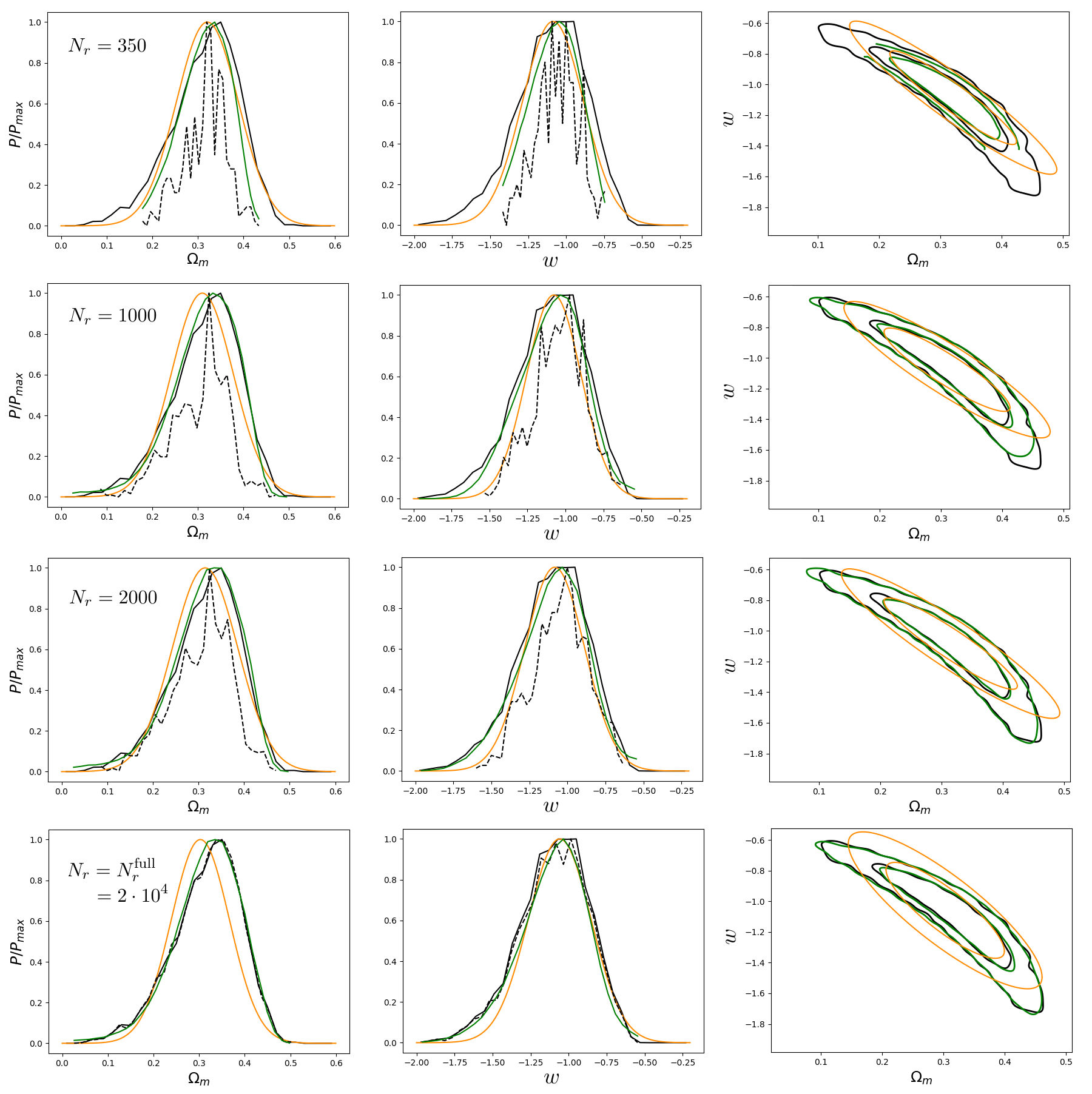}
\caption{\label{fig:fig4} In each row of this figure we show the marginalized one-dimensional posterior distributions for the normalized current matter density parameter $\Omega_m$ (left column) and the dark energy EoS parameter $w$ (middle column), together with the $1\sigma$ and $2\sigma$ confidence contours in the $(\Omega_m,w)$-plane (right column), all obtained from the analysis of the SnIa data from the Pantheon+MCT compilation (Scolnic et al. 2018, Riess et al. 2018). We plot: the posterior obtained from the usual MonteCarlo with $N^{\rm full}_r=2\cdot 10^4$ sampling points (solid black lines); the posterior obtained also with the traditional MonteCarlo, but using a number of sampling points $N_r\leq N^{\rm full}_r$, as indicated in each row (dashed black lines); the Gaussian fit \eqref{eq:fitGaussianImproved} to the latter (in orange); and the non-Gaussian fit \eqref{eq:full-fit} (in green). See the main text in Sec. \ref{sec:SNIa} for further details.}
\end{figure*}

Other algorithms can be implemented to force the marginalization procedure to perform the additional sampling only within the good, well-sampled region. For instance, the marginalization sampling could be performed  along segments connecting nearby points of the first sampling; this and other alternative techniques will be explored in future work.


\section{Application: Supernovae Ia}\label{sec:SNIa}

\begin{figure*}
\includegraphics[width=6.4in, height=5.85in]{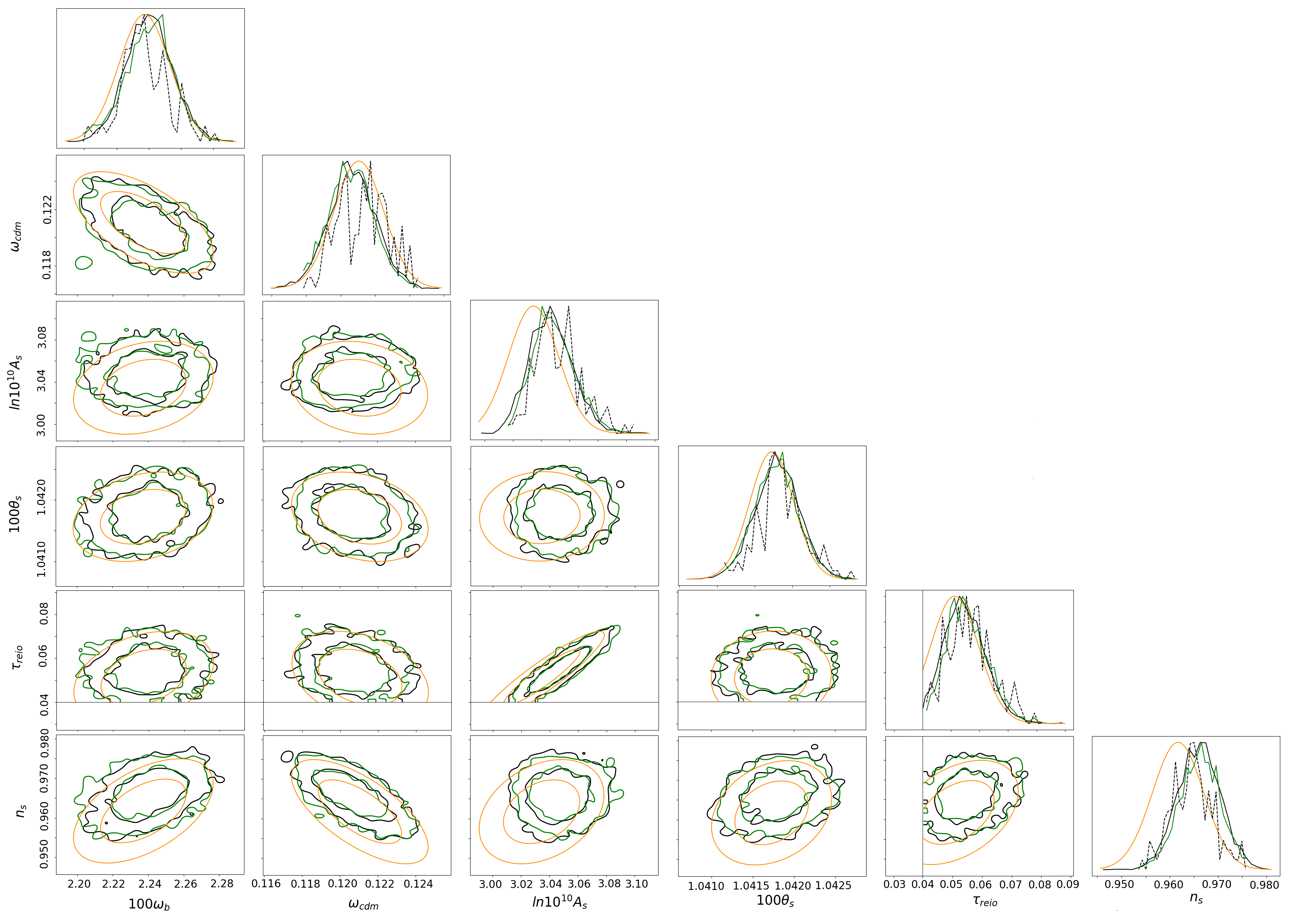}
\caption{\label{fig:fig5} As in Fig. \ref{fig:fig4}, but for the six $\Lambda$CDM parameters $(\omega_b,\omega_{cdm},\ln(10^{10}A_s),100\theta_s,\tau_{reio},n_s)$. These constraints have been obtained using the Planck 2018 TT,TE,EE+lowE likelihood (Planck Collaboration 2018), and carrying out the corresponding marginalizations, see Sec. \ref{sec:CMB} for details. We show: the posterior obtained from the usual MonteCarlo with $N^{\rm full}_r= 3\cdot 10^5$ sampling points (solid black lines); the same, but using a number of sampling points $N_r=1.3\cdot 10^4$ (dashed black lines, only shown on the 1D plots); the Gaussian fit \eqref{eq:fitGaussianImproved} using the latter (in orange); and the non-Gaussian fit \eqref{eq:full-fit} (in green), up to cubic order.}
\end{figure*} 

As a first cosmological application of the MonteCarlo Posterior Fit explained in the preceding sections, we study its performance in the case in which the $w$CDM parametrization (Turner \& White 1997) of the dark energy (DE) equation of state (EoS) is fitted to the data of SnIa from the Pantheon+MCT compilation (Scolnic et al. 2018, Riess et al. 2018). We use the compressed likelihood built from the original distance moduli, which constrains the Hubble rate $E(z)=H(z)/H_0$ at six different redshifts, see Table 6 and also Fig. 3 of (Riess et al. 2018). The only two cosmological parameters that enter the fit are the normalized matter density, $\Omega_m$, and the constant DE EoS parameter, $w$. We have obtained a MonteCarlo Markov chain containing $N_r^{\rm full}=2\cdot 10^4$ points using the MC sampler \texttt{MontePython}\footnote{\textcolor{blue}{\url{http://baudren.github.io/montepython.html}}} (Audren et al. 2013), and computing $E(z)$ with the Einstein-Boltzmann system solver \texttt{CLASS}\footnote{\textcolor{blue}{\url{http://lesgourg.github.io/class public/class.html}}} (Blas, Lesgourgues \& Tram 2011). Using $N_r^{\rm full}$, the Gelman-Rubin diagnostic (Gelman \& Rubin 1992) for the two parameters reads $\hat{R}-1\sim 10^{-3}$, which indicates convergence. Then we have split the full chain in subsamples containing $N_r<N_r^{\rm full}$ points in order to fit them with both, the Gaussian \eqref{eq:fitGaussianImproved} and the non-Gaussian \eqref{eq:full-fit} distributions. As explained earlier, the latter introduces cubic and fourth order corrections to the pure Gaussian fit. Once the second-parameters are computed following the procedure explained in Secs. \ref{sec:G} and \ref{sec:NG}, we have to carry out the marginalization over $\Omega_m$ and $w$ in order to obtain the 1D posteriors. This is analytical in the case of \eqref{eq:fitGaussianImproved} (at least for positive-definite matrices), but for the non-Gaussian distribution we need to carry out a marginalization MC, as described in Sec. \ref{sec:Marginalization}. Then we can build the corresponding histograms, from which one can later on draw not only the 1D distributions but also the contour plots in the $(\Omega_m,w)$-plane.  

\begin{figure*}
\includegraphics[width=6in, height=5in]{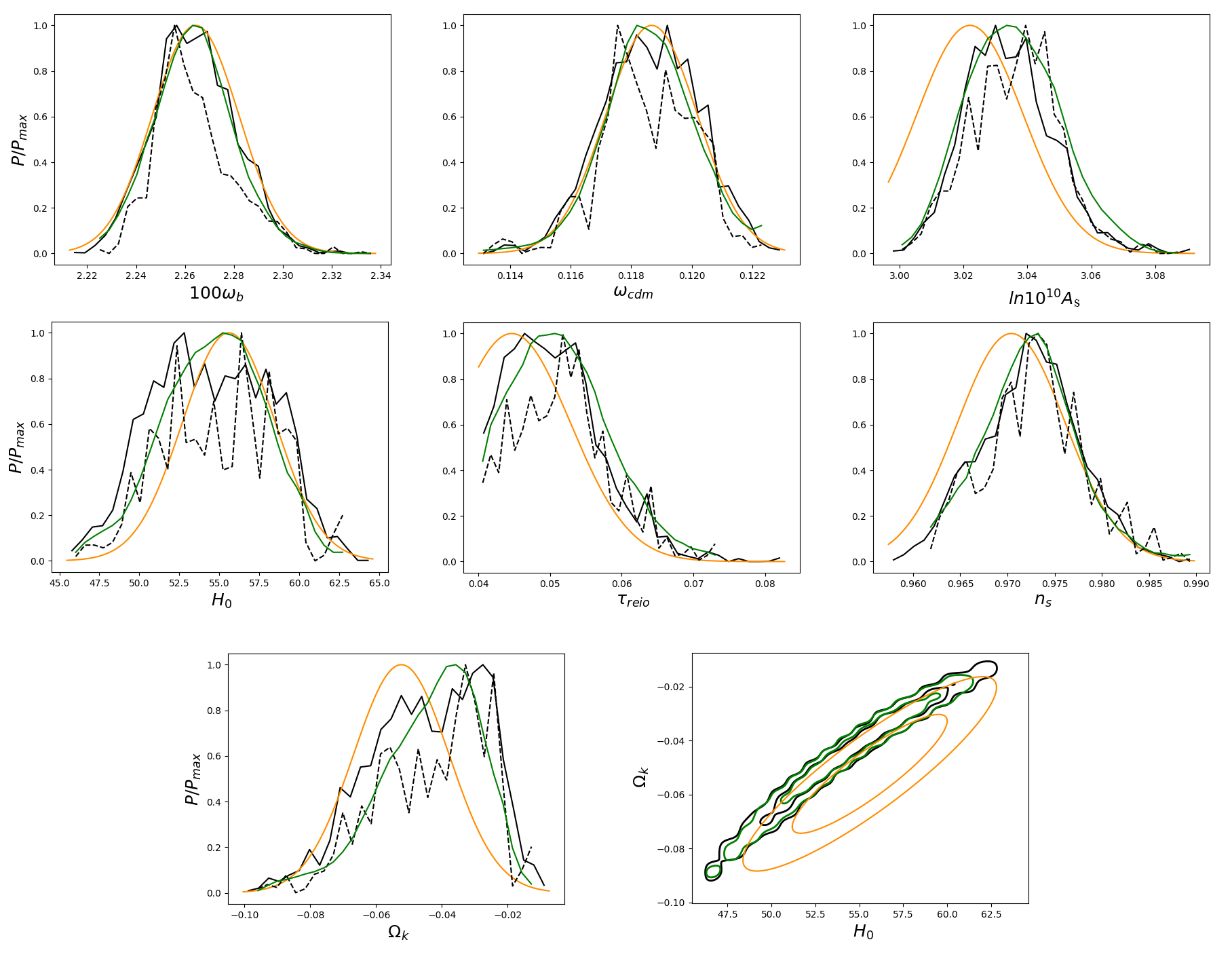}
\caption{\label{fig:OmegakPlot} 1D exact and reconstructed posteriors obtained for the seven parameters $(\omega_b,\omega_{cdm},\ln(10^{10}A_s),H_0,\tau_{reio},n_s,\Omega_k)$ of the $\Lambda$CDM model with spatial curvature. $H_0$ is given is km/s/Mpc. In the right plot of the last row, we also present the $1\sigma$ and $2\sigma$ 2D contours in the $(H_0,\Omega_k)$-plane. These correspond to the constraints obtained using the Planck 2018 TT,TE,EE+lowE likelihood (Planck Collaboration 2018), cf. Fig. 26 therein. We show: the posterior obtained from the usual MonteCarlo with $N_r= 7\cdot 10^5$ sampling points (solid black lines); the same, but using $N_r=1.7\cdot 10^5$ (dashed black lines); the Gaussian fit \eqref{eq:fitGaussianImproved} using the latter (in orange); and the non-Gaussian fit \eqref{eq:full-fit} (in green). We include all the 28 parameters in the $C,P,M$ terms, but only the seven cosmological parameters plus 10 nuisance parameters in the $S,K$ terms. See the main text for further details.}
\end{figure*}

Our results are shown in Fig. \ref{fig:fig4}. Some comments are in order. First, one can see therein that the distribution in the  $(\Omega_m,w)$-plane is affected by clear non-Gaussianities, which can be duly quantified using Eq. \eqref{eq:convCriterion}, giving $\tau=1.11$ (cf. Appendix \ref{sec:AppendixB} for details). Second, the ability of the method to describe the underlying distribution is really good, even when a very small number of points is employed in the fitting analysis needed to extract the second-parameters. For instance, from the first two rows of Fig. \ref{fig:fig4} it is clear that even employing $N_r=350$ or $N_r=1000$ (corresponding to fractions of $\sim 2\%$ and $\sim 5\%$ of the full sample, respectively) we obtain results that basically match with those obtained from $N_r^{\rm full}$ sampling points. The differences in the means for the two parameters are lower than $0.05\sigma$ when we employ $N_r=1000$. The convergence of the method is quite fast. Third, as we have mentioned earlier, in general we can rely on the good description of the underlying distribution only in those regions of parameter space containing sampling points. This is why for the non-Gaussian fit we only show the results in these regions.

Of course, here we have generated $N_r^{\rm full}$ sampling points just for illustrative purposes. In a real situation we want to make use of the MonteCarlo Posterior Fit to save computational time, so it is convenient to define a criterion that allows us to stop the MC sampling once the fitting distribution \eqref{eq:fitGaussianImproved} remains stable under the addition of more sampling points. This will mean that it has already converged to the final result. Many criteria are possible. For instance, one can apply the method repeatedly in parallel to the main MonteCarlo and then compute e.g. the maximum and the $68.3\%$ c.l. intervals of the marginalized 1D distributions for the desired parameters using \eqref{eq:fitGaussianImproved}. One can decide to stop the main MonteCarlo once the relative differences between the values obtained for these quantities in two consecutive evaluations is below a certain threshold, say e.g. $5\%$. This criterion is not very efficient, though, since it requires the marginalization of the fitted distribution. For more practical criteria which do not require any marginalization step cf. the Appendix \ref{sec:AppendixC} and the results of Sec. \ref{sec:CMB}.

The advantages of the MonteCarlo Posterior Fit will become more evident in the next section, where we use CMB data and of course consider a much bigger parameter space. These facts clearly increase the complexity of the problem. The evaluation of the exact SnIa likelihood is already quite fast, and the absolute gain in computational time due to the use of \eqref{eq:fitGaussianImproved} is not very substantial. This example, though, has served us to nicely illustrate the ability of our fit to capture very precisely the non-Gaussian features of the underlying distribution even with a small sampling, and to show that the convergence is also quite fast.

\section{Application: CMB}\label{sec:CMB}

We start applying the MonteCarlo Posterior Fit to the chain obtained for the $\Lambda$CDM and using the Planck 2018 TT,TE,EE+lowE likelihood (Planck Collaboration 2018). In this case we have the 6 usual cosmological parameters of the standard model, i.e. ($\omega_b$, $\omega_{cdm}$, $\ln(10^{10}A_s)$, $100\theta_s$, $\tau_{reio}$, $n_s$), and also the 21 nuisance parameters that enter Planck's likelihood. Thus, we deal with a parameter space of $N=27$ dimensions. This means that the square matrix $A$ \eqref{eq:A_matrix} has now $31465\times 31465$ elements. It is difficult (very demanding from the computational point of view) to perform the inversion of this matrix, something that is needed to obtain the second-parameters through \eqref{eq:solng}. In order to reduce the size of $A$ we can opt, as mentioned before, to consider e.g. only the elements of the matrix $K$ associated to the main cosmological parameters, or just stick to the cubic correction of the Gaussian fit. Later on we will analyze another example considering some elements of the $K$ matrix for illustrative purposes, but in the current case we just consider the third order correction, since it already provides excellent results, as we explicitly show in Fig. \ref{fig:fig5}. 
\begin{figure*}
\includegraphics[width=6in, height=2in]{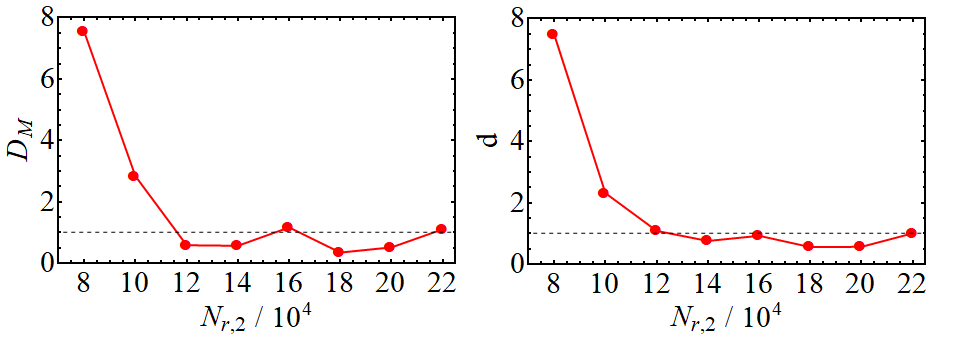}
\caption{\label{fig:criteria} Convergence criteria as a function of the number of sampled points. We can see that the Mahalanobis distance $D_M$ \eqref{eq:Maha} and the criterion $d$ \eqref{eq:distMConv} are below $1$ for $N_{r,2}\gtrsim 12\cdot 10^{4}$, which indicates that for such number of sampling points we can already stop the first MonteCarlo and perform the marginalization of the fitted distribution. We have chosen here $f=0.2$ and $\tilde{f}=0.04$, cf. Appendix \ref{sec:AppendixC}. The values of the criteria at the locations marked by the dots have been computed using the numbers of $N_{r,2}$ indicated in the $x$-axis and $N_{r,1}=N_{r,2}-2\cdot 10^{4}$.}
\end{figure*}
To obtain such plot we have basically proceeded as in the preceding section. Now we have generated a full chain of $N_r^{\rm full}=3\cdot 10^{5}$ sampling points, from which we can draw 1D posteriors and 2D contour plots which are very close to the ones of the exact underlying distribution, since $\hat{R}-1<10^{-2}$ for all the parameters. These results correspond to the solid black lines in Fig. \ref{fig:fig5}. On the other hand, we have taken a subchain containing only  $4.3\%$ of the points contained in the full sample, i.e. $N_r=1.3\cdot 10^{4}$ points, and we have applied the Gaussian and non-Gaussian fits (including the third order correction through the matrix $S$). We have plotted the output in orange and green, respectively. One can see that in this case the Gaussian fit already provides pretty good results, e.g. the peaks of the one-dimensional marginalized fitted distributions are very close to the exact ones, being the difference in all cases lower or much lower than $1\sigma$ depending on the parameter. Of course, this will not necessarily hold in general, so one needs to check how much improvement there is with the higher-order terms. In this case, we find that  the improvement introduced by the non-Gaussian corrections is quite remarkable. The corresponding results are almost indistinguishable from the ``exact'' ones, obtained from the full chain with $N_r^{\rm full}$ points. We could have produced a larger sample, taking advantage of the efficiency of our method, but here we opt to generate $N_r^{\rm full}$ to compare our results with the standard approach on equal levels of sampling noise. The results are much better than the ones obtained from the usual MonteCarlo with $N_r$ points, which are plotted with dashed black lines. The relative gain in terms of computational time offered by the MC Posterior Fit method when compared with the usual approach is also important. The former is $\sim 10$ times faster, due to the fact that making use of the fitted distribution one can skip the expensive calculation of the Einstein-Boltzmann equations with \texttt{CLASS} in the marginalization MC, something that cannot be avoided in the standard method. The improvement in efficiency provided by the MC Posterior Fit depends also on the exact form of the fitting distribution under consideration and other technical aspects, as e.g. the typical length of the steps in the marginalization MC. In this example we have used as proposal distribution a multivariate Gaussian with the covariance matrix computed from the $N_r$ sampling points of the original MC, but without shortening the steps. Many points are generated outside the $N$-dimensional box described in Sec. \ref{sec:Marginalization}, which slows the routine a little bit down. Further gain in efficiency can be obtained by shortening the steps, especially when the level of non-Gaussianity is not small, as in the next example.

We also apply the method to an even  more complex case, with one additional parameter and a higher degree of non-Gaussianity: the $\Lambda$CDM model with spatial curvature, again under the Planck 2018 TT,TE,EE+lowE likelihood (Planck Collaboration 2018). The results are shown in Fig. \ref{fig:OmegakPlot}. For the non-Gaussian fit we include all the 28 parameters in the $C,P,M$ terms, but only the seven cosmological parameters plus 10 nuisance parameters in the $S,K$ terms of Eq. \eqref{eq:full-fit}\footnote{We have explicitly checked that the choice of the set of 10 nuisance parameters included in the $S,K$ terms is not important in this case. The results remain completely stable under different combinations.}. We employ $N_r=1.7\cdot 10^{5}$ sampling points in the first MC. In Fig. \ref{fig:criteria} we show that with this number of sampling points we fulfill the convergence criteria explained in Appendix \ref{sec:AppendixC}, which means that we can already perform a good reconstruction of the underlying distribution with the MC Posterior Fit, see the comments in the caption. Then we compute the second-parameters and subsequently obtain one million points through the marginalization MC. This is done quite efficiently, since one single evaluation of $L_M$ is $\sim 40$ times faster than the evaluation of $L$.  Here we have employed shorter steps in the Metropolis-Hastings algorithm of the marginalization MC, by dividing the standard deviations of all the parameters in the covariance matrix of the proposal multivariate Gaussian by 10. The results are excellent. We have checked that the means and corresponding uncertainties match with those provided by the Planck collaboration \footnote{See Sec. 15.3 of \textcolor{blue}{\url{https://wiki.cosmos.esa.int/planck-legacy-archive/images/4/43/Baseline\_params\_table\_2018\_68pc\_v2.pdf}}}. For instance, for the curvature and Hubble parameters we obtain $\Omega_k=-0.044^{+0.017}_{-0.012}$ and $H_0=(54.6^{+3.3}_{-3.1})$ km/s/Mpc, respectively, whereas Planck report $\Omega_k=-0.044^{+0.018}_{-0.015}$ and $H_0=(54.4^{+3.3}_{-4.0})$ km/s/Mpc. They are fully compatible. To obtain a comparable degree of accuracy with the standard approach we would have needed a total number of $N_r^{\rm full}\gtrsim 10^6$ sampling points obtained with \texttt{CLASS}+\texttt{MontePython}. The overall computational time with the MC Posterior Fit is reduced by a factor $\sim 6$ w.r.t. the conventional method, and a factor $\sim 8$ if we stop the first MC a little bit earlier, when $N_r\sim 1.2\cdot 10^{5}$ (cf. again Fig. \ref{fig:criteria}).


\section{Improving upon the Fisher matrix}

\begin{figure*}
\includegraphics[width=5in, height=3in]{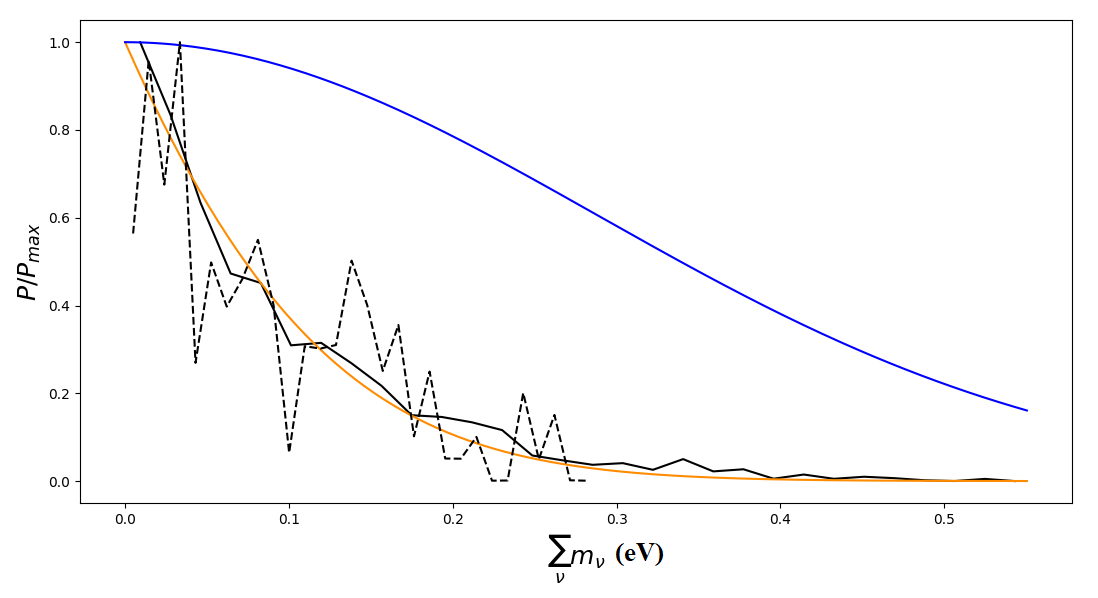}
\caption{\label{fig:mnu} Reconstruction of the 1D posterior of the sum of the neutrino masses obtained from the analysis of the Planck 2018 TT,TE,EE+lowE likelihood (Planck Collaboration 2018), assuming the $\Lambda$CDM and considering as in that reference the case of three degenerate neutrinos with equal mass. We have of course marginalized over the 6 $\Lambda$CDM parameters and 21 nuisance parameters of the Planck likelihood. The various curves represent: the posterior obtained from the standard MonteCarlo with $N_{r,1}= 3.6\cdot 10^5$ and $N_{r,2}=2\cdot 10^4$ sampling points (solid and dashed black lines, respectively); the Gaussian fit \eqref{eq:fitGaussianImproved} to the $N_{r,2}$ points (in orange); the Fisher matrix approximation (in blue).}
\end{figure*}

As mentioned in the Introduction, the formalism above can be used also in another context, namely when we want to  forecast the performance of future observations assuming a specific theoretical model (e.g. $\Lambda$CDM). In this case we know from the start the vector of best fit (maximum likelihood) parameters $\hat\theta_\alpha$.
Then, given a posterior that depends on $N$ parameters $\theta_{\alpha},$
the usual Fisher matrix  is obtained as
\begin{equation}
F_{\alpha\beta}=-\frac{\partial^{2}\log L}{\partial\theta_{\alpha}\partial\theta_{\beta}}\Biggr\rvert_{\mathbf{\theta}=\hat{\mathbf{\theta}}}\,.
\end{equation}
The FM is essentially a simple way to approximate
a generic likelihood with a Gaussian near its maximum
\begin{equation}\label{eq:fitFisher}
L_{F}(\theta)=\hat{L}\exp\left[-\frac{1}{2}(\theta-\hat{\theta})_{\alpha}F_{\alpha\beta}(\theta-\hat{\theta})_{\beta}\right]\,,
\end{equation}
where $\hat{L}=L(\hat{\theta})$ is the maximum of the likelihood. Deviations from Gaussianity can be modeled by extending to higher-order derivatives, as proposed in (Sellentin 2014, Sellentin, Quartin \& Amendola 2015).
The great advantage of the Fisher matrix is that it just requires to evaluate the posterior at a handful of points (of the order of the square of the number of parameters)  near the best fit values, in order to evaluate the Hessian $F_{\alpha\beta}$. 
However, as already mentioned, in this way the approximation is only good near the maximum (and only if a well-defined local maximum with zero first derivatives exists)
and can be quite bad as soon as one moves away from it in parameter
space. The full exploration of the posterior through a MonteCarlo Markov Chain, as already mentioned, might however be very demanding. Our method tries to find an optimal balance between these two extremes.

The application to the forecast problem is very much the same as we have already seen. The main difference is that now one needs to generate a mock dataset assuming a fiducial model in order to build the likelihood. Then one generates a number of points in parameter space, either by employing a MCMC algorithm or by any other way of sampling the space (e.g. through a regular or irregular grid if $N$ is low enough) and evaluates the posterior corresponding to those points. Then one adopts either the simplest Gaussian form \eqref{eq:fitGaussian}, or the improved Gaussian \eqref{eq:fitGaussianImproved}, or finally the non-Gaussian one \eqref{eq:full-fit}, depending on the degree of accuracy required, and finds the  second-parameters using e.g. Eq. (\ref{eq:solng}). Even when using the Gaussian form, one is advised to keep the linear terms in $C$ and $P$ in order to deal with cases in which the posterior does not have a maximum with flat first derivatives (perhaps because the maximum happens to lie on the border of a sharp prior). 

In Figure \ref{fig:mnu}, we compare the performance of the standard FM with the MonteCarlo Posterior Gaussian fit in a concrete example, in which we reconstruct the one-dimensional posterior distribution of the sum of the neutrino masses, i.e. $\sum_\nu m_\nu$, obtained from the analysis of the Planck 2018 TT,TE,EE+ lowE likelihood (Planck Collaboration 2018). We assume the non-minimal extension of the $\Lambda$CDM model and consider $\sum_\nu m_\nu>0$ as an additional cosmological parameter. We study the case of three degenerate neutrinos with equal mass, as in (Planck Collaboration 2018). In this example the advantage offered by the MC Posterior Fit method is evident, and one could easily think of cases in which the standard FM fails arbitrarily badly, for instance a posterior such that the standard Fisher matrix is singular.

Let us recap the advantages
of the MonteCarlo Posterior Fit over the FM:
1) The FM is by definition accurate only near the peak of the posterior;
the MC Posterior Fit depends on a larger sampling of the distribution, so it is able to describe much better the underlying distribution far away from the peak, especially in its non-Gaussian implementation;
2) The posterior might even lack a  maximum with vanishing first derivatives. In this case the FM
will be totally unreliable. The MC Posterior Fit is instead stable (see Fig. \ref{fig:mnu});
3) If there are strongly degenerate directions, the FM will be almost
singular. The MC Posterior Fit does not have this problem.


\section{Conclusions}

In this paper we presented a novel method, called Montecarlo Posterior Fit, which improves the efficiency
of the sampling of likelihoods offered by the usual MC approaches. We use a multivariate non-Gaussian function to fit a much smaller number of sampling points 
than the one needed to reconstruct the posterior with the usual 
methods. The second-parameters that characterize the fitting function can be easily computed by solving a simple linear 
system of equations, and then a marginalization MC is needed to sample the fitted distribution and to subsequently obtain 
the 1D posteriors and 2D confidence regions for the various parameters of the model under study. The advantage of this method becomes important
in those cases in which the evaluation of the fitted distribution is much faster than the original likelihood. This is clearly 
the case e.g. when the use of Einstein-Boltzmann codes  is required to compute the observables that enter the likelihood. 

We have discussed some examples 
of applications based on data on supernovae of Type Ia and CMB, which clearly illustrate the power of the method. In the CMB case, the method is able to 
reduce the computational time by a factor from six to ten. The MonteCarlo Posterior Fit can be 
also very useful to boost the efficiency of Bayesian analyses that involve $N$-body simulations. In this case the reduction in computational time can 
be even more substantial. This deserves a dedicated study, which we leave for a future work.

Another very interesting application of our method is found in the elaboration of forecasts. In order to avoid the lengthy exploration of the whole parameter space one usually
makes use of the Fisher matrix method, which approximates the exact likelihood by a multivariate Gaussian and requires only the estimation of its second derivatives
at the maximum. This requires only few evaluations of the likelihood around the peak. As we have discussed in this paper, the Fisher matrix has some important drawbacks, though. For instance, 
it only describes correctly the underlying distribution near the peak, and can fail once we depart from it; or can lead to very bad results in the case in which the exact likelihood
has no flat maximum, for instance because truncated by the prior boundaries. The MC Posterior Fit can correct all these deficiencies of the Fisher matrix method, bridging the gap between it and the time-consuming full MonteCarlo approach.

\section*{Acknowledgements}
AGV is funded by the Deutsche Forschungsgemeinschaft (DFG) - Project number 415335479. We thank Viviana Niro for her help in the configuration of \texttt{MontePython} in the case studied in Fig. \ref{fig:mnu}.

\section*{Data availability}
We provide the references of all the data sources in the article. The code in which we have implemented the MonteCarlo Posterior Fit method, \texttt{MCPostFit}, is publicly available at: \textcolor{blue}{\url{https://github.com/adriagova/MCPostFit}}.

\appendix 

\section{Correcting the peak's location with Newton-Raphson}\label{sec:AppendixA}

It is also possible to apply an iterative routine in order to look for the correction of the position and height of the peak of our fitting distribution. For instance, one based on the Newton-Raphson method (see e.g. Press et al. 2007). Let us focus at the moment on the pure Gaussian fit, i.e. let us consider 
\begin{equation}
L_M(\theta)=\bar{L}\exp\left[-\frac{1}{2}(C+(\theta-\tilde{\theta})_\alpha M_{\alpha\beta}(\theta-\tilde{\theta})_\beta)\right]\,.
\end{equation}

\begin{figure*}
\includegraphics[width=18cm]{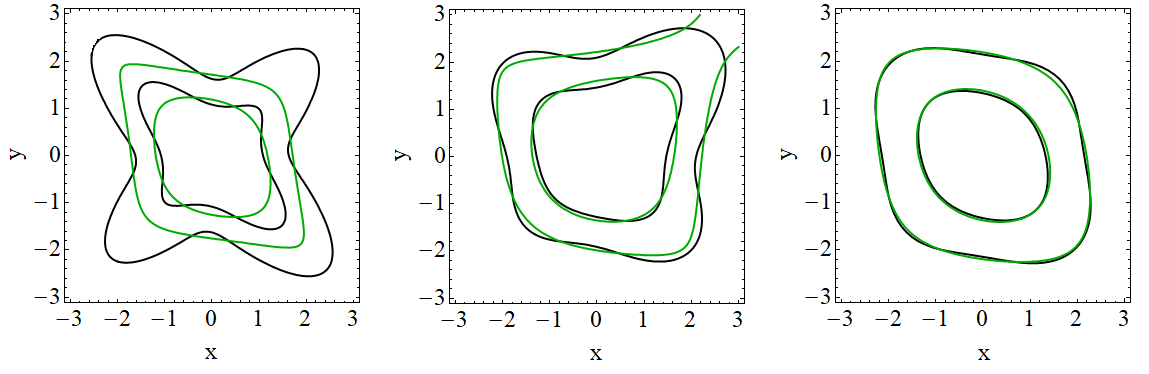}
\caption{$1\sigma$ and $2\sigma$ confidence contours associated to the analytical posteriors \eqref{eq:starry}-\eqref{eq:mix} (in black) and their reconstructed shapes obtained with the MC Posterior Fit method (in green), using the quartic non-Gaussian fitting distribution \eqref{eq:full-fit}. See the comments in Appendix \ref{sec:AppendixExamples}.\label{fig:figAppendixB}}  
\end{figure*}

In this case, one can start computing \eqref{eq:resultM} at a given $\tilde{\theta}^{(1)}$ (for instance, using the vector $\theta$ with largest $L$ in the Markov chain) to obtain the initial matrix $M^{(1)}$ (notice that we can take $C^{(1)}=0$), and then solve iteratively the following system of equations until the cost function \eqref{eq:chi2} is minimized,
\begin{multline}
\frac{\partial\chi^2_M}{\partial\tilde{\theta}_\alpha} \Bigg\rvert_{n}
+\frac{\partial^2 \chi^2_M}{\partial\tilde{\theta}_\alpha\partial\tilde{\theta}_\beta} \Biggr\rvert_{n}(\tilde{\theta}_\beta^{(n+1)}-\tilde{\theta}_\beta^{(n)})
\\+\frac{\partial^2 \chi^2_M}{\partial\tilde{\theta}_\alpha\partial M_{\beta \mu}} \Biggr\rvert_{n}(M_{\beta\mu}^{(n+1)}-M_{\beta\mu}^{(n)})+\frac{\partial^2 \chi^2_M}{\partial \tilde{\theta}_\alpha\partial C} \Biggr\rvert_{n}(C^{(n+1)}-C^{(n)})=0\,,
\end{multline}
\begin{multline}
\frac{\partial\chi^2_M}{\partial C} \Bigg\rvert_{n}
+\frac{\partial^2 \chi^2_M}{\partial C^2} \Biggr\rvert_{n}(C^{(n+1)}-C^{(n)})\\
+\frac{\partial^2 \chi^2_M}{\partial C\partial M_{\mu\nu}} \Biggr\rvert_{n}(M_{\mu\nu}^{(n+1)}-M_{\mu\nu}^{(n)})+\frac{\partial^2 \chi^2_M}{\partial \tilde{\theta}_\mu\partial C} \Biggr\rvert_{n}(\tilde{\theta}_\mu^{(n+1)}-\tilde{\theta}_\mu^{(n)})=0\,,
\end{multline}
\begin{multline}
\frac{\partial\chi^2_M}{\partial M_{\alpha\beta}} \Biggr\rvert_{n} +\frac{\partial^2 \chi^2_M}{\partial M_{\alpha\beta}\partial M_{\mu\nu}} \Biggr\rvert_{n}(M_{\mu\nu}^{(n+1)}-M_{\mu\nu}^{(n)}) \\
+\frac{\partial^2 \chi^2_M}{\partial\tilde{\theta}_\mu\partial M_{\alpha\beta}} \Biggr\rvert_{n}(\tilde{\theta}_\mu^{(n+1)}-\tilde{\theta}_\mu^{(n)})+\frac{\partial^2 \chi^2_M}{\partial C\partial M_{\alpha\beta}} \Biggr\rvert_{n}(C^{(n+1)}-C^{(n)})=0\,,
\end{multline}
where all the derivatives are evaluated at $(\tilde{\theta}^{(n)},C^{(n)},M^{(n)})$. This system can be written in the standard Newton-Raphson form by using again the compact notation $M_{\mu\nu}\to M_{a}$, as in Secs. \ref{sec:G} and \ref{sec:NG}, and building a vector of $d=N+1+N(N+1)/2=1+N(N+3)/2$ dimensions, $\vec{\xi}=(\tilde{\theta}_\alpha,C,M_{a})$. The result reads,
\begin{equation}\label{eq:NRmethod}
\vec{\xi}^{(n+1)} = \vec{\xi}^{(n)}-\mathcal{H}_\xi^{-1}\vec{\nabla}_\xi\chi^2_M\Biggr\rvert_{\vec{\xi}^{(n)}}\,,
\end{equation}
with $\vec{\nabla}_\xi$ and $\mathcal{H}_\xi$ being the gradient and the Hessian matrix associated to $\vec{\xi}$, respectively. It is important to recall that the original Newton-Raphson method does not always lead to a minimum. It can also lead to a saddle point or even produce divergent results when the Hessian is not positive definite. It is possible to modify slightly this method to solve this issue. We have done so by implementing the so-called Levenberg-Marquardt correction (Levenberg 1944, Marquardt 1963), which basically consists on susbtituting the Hessian by $\mathcal{H}\to\mathcal{H}+\mu\, I_{d\times d}$ (with $I_{d\times d}$ being the identity matrix in $d$ dimensions) and changing properly $\mu$ at each iteration step to ensure the positive definiteness of the resulting matrix. We can stop the iterative process e.g. when the relative change $|\xi_i^{(n+1)}/\xi_i^{(n)}-1|$ is below a desired value, which can be different for every $i$. The lowest is this value, the highest the precision we demand in our calculation.

A straightforward generalization of this method can also be employed for the non-Gaussian fit \eqref{eq:full-fit} (cf. Sec. \ref{sec:NG}), considering equation \eqref{eq:NRmethod} and working now with the vector $\vec{\xi}=(\tilde{\theta}_\alpha,C,M_{a},S_{\bar{a}},K_{\tilde{a}})$ and the corresponding gradient and Hessian. This iterative approach (and more concretely, the calculation of the Hessian matrix) can be, though, quite expensive in terms of computational time, specially if we are working with a high-dimensional parameter space, as in Sec. \ref{sec:CMB}. The procedure we have explained in the main body of the paper corrects the position of the peak in a much more efficient way. We have implemented also the alternative iterative method just to cross-check the results that we have presented in the paper. We have verified that both methods lead to the same results, as expected. 


\section{Fit to starry/boxy posteriors}\label{sec:AppendixExamples}

In this appendix we study the ability of the quartic non-Gaussian distribution \eqref{eq:full-fit} to fit posteriors with starry and boxy shapes, which are not encountered in the cosmological applications of Secs. \ref{sec:SNIa} and \ref{sec:CMB}. We are not concerned here with the MC optimization, but just with shape flexibility.  For the sake of simplicity we stick  to two-dimensional parameter spaces and consider the three following analytical posterior distributions obtained from the combination of multivariate Gaussians:

\begin{multline}\label{eq:starry}
h(x,y) =  0.088\,e^{-2.78\left(\frac{x^2}{2} - 0.8xy+\frac{y^2}{2}\right)}+\\ 
      +0.177\,e^{-2.78\left(\frac{x^2}{2} + 0.8xy + \frac{y^2}{2}\right)}
\end{multline}

\begin{multline}\label{eq:boxy}
f(x,y) =  0.099\, e^{-1.56\left(\frac{(x-0.5)^2}{2} - 0.6(x-0.5)(y-0.5)+\frac{(y-0.5)^2}{2}\right)}+\\ 
      +0.099\,e^{-1.56\left(\frac{x^2}{2} + 0.6xy + \frac{y^2}{2}\right)}
\end{multline}

\begin{multline}\label{eq:mix}
g(x,y) =  0.058\, e^{-1.19\left(\frac{x^2}{2} - 0.4xy+\frac{y^2}{2}\right)}+\\ 
      +0.123\,e^{-1.33\left(\frac{x^2}{2} + 0.5xy + \frac{y^2}{2}\right)}
\end{multline}

The results are shown in Fig. \ref{fig:figAppendixB}. The black contours are obtained from the exact distributions \eqref{eq:starry}-\eqref{eq:mix} and the green ones from the corresponding posterior non-Gaussian fits \eqref{eq:full-fit}. The leftmost plot clearly shows that if the posterior is extremely starry, then Eq.  \eqref{eq:full-fit} is unable to provide an accurate description, capturing only some of its features. If the degree of "starriness" of the underlying distribution is lower, with more boxy contours, then our method seems to work very well, as evident from the second and third plots of Fig. \ref{fig:figAppendixB}.


\section{Assessing the degree of non-Gaussianity}\label{sec:AppendixB}

One can assess whether the non-Gaussian terms are important 
by evaluating  the covariance matrix with and
without the non-Gaussian correction, again focusing only on the fourth-order
term. 
We discuss this test briefly here as a base for further study, although we only used it in the SnIa application discussed in Sec. \ref{sec:SNIa}.

Assuming the deviation from Gaussianity is not excessive and considering in \eqref{eq:full-fit} the $S$- and $K$-terms associated to all the parameters, we can build the following approximation
\begin{multline}
\langle(\theta-\hat{\theta})_{\mu}(\theta-\hat{\theta})_{\nu}\rangle =\\
=\tilde{L}\int d^{N}\theta D_{\mu\nu}\exp\left[-\frac{1}{2}M_{\alpha\beta}D_{\alpha\beta}-\frac{1}{2}S_{\alpha\beta\gamma}D_{\alpha\beta\gamma}-\frac{1}{2}K_{\alpha\beta\gamma\delta}D_{\alpha\beta\gamma\delta}\right]\\
\approx \tilde{L}\int d^{N}\theta D_{\mu\nu}\exp\left[-\frac{1}{2}M_{\alpha\beta}D_{\alpha\beta}\right]\\\phantom{XXXX}\times \left(1-\frac{1}{2}S_{\alpha\beta\gamma}D_{\alpha\beta\gamma}-\frac{1}{2}K_{\alpha\beta\gamma\delta}D_{\alpha\beta\gamma\delta}\right)\\
 =M_{\mu\nu}^{-1}-\frac{1}{2}K_{\alpha\beta\gamma\delta}\tilde{L}\int d^{N}\theta D_{\mu\nu}D_{\alpha\beta\gamma\delta}\exp\left[-\frac{1}{2}M_{\alpha\beta}D_{\alpha\beta}\right]\\
 =M_{\mu\nu}^{-1}+\kappa_{\mu\nu}\,,
\end{multline}
where
\begin{equation}
\kappa_{\mu\nu}=-\frac{1}{2}K_{\alpha\beta\gamma\delta}(M_{\mu\nu}^{-1}M_{\alpha\beta}^{-1}M_{\gamma\delta}^{-1}+14\mathrm{\,distinct\:permutations})\,.
\end{equation}
Then we can finally compactify the deviation into a single number,
\begin{equation}\label{eq:convCriterion}
\tau=|\det (M\kappa)|\,,
\end{equation}
with the matrix inside the determinant taking the following form:
\begin{equation}
(M\kappa)_{\mu\nu}=-\frac{3}{2}\delta_{\mu\nu}K_{\alpha\beta\gamma\theta}M^{-1}_{\alpha\beta}M^{-1}_{\gamma\theta}-6 K_{\mu\beta\gamma\theta}M^{-1}_{\nu\beta}M^{-1}_{\gamma\theta}\,.  
\end{equation}
Although clearly a single number cannot fully characterize the degree of multi-dimensional non-Gaussianity, we can expect that  $\tau \ge 1$ signals a significant non-Gaussianity of the fitted distribution. 


\section{Criteria to stop the first MC}\label{sec:AppendixC}

We suggest two convergence criteria that can be used  to decide whether we have reached enough sampling points $N_r$. Both make use of the Gaussian fit \eqref{eq:fitGaussianImproved} and measure the stability of the mean and the covariance matrix under the increase of $N_r$. Imagine we want to compare the situation when we have $N_{r,1}$ and $N_{r,2}>N_{r,1}$ sampling points. The fitted Gaussian obtained from these samples will be centered at $\hat{\theta}_{s1}$ and $\hat{\theta}_{s2}$ and will have associated matrices $M_{s1}$ and $M_{s2}$, respectively. In order to evaluate the stability of the mean values, we can compute the so-called Mahalanobis distance (Mahalanobis 1936) between them, $D_M(\hat{\theta}_{s1},\hat{\theta}_{s2})$, using $M_{s2}$ for the metric of the parameter space. It is given by the following simple formula,
\begin{equation}\label{eq:Maha}
D_M(\hat{\theta}_{s1},\hat{\theta}_{s2}) = \sqrt{(\hat{\theta}_{s1}-\hat{\theta}_{s2})^T M_{s2}(\hat{\theta}_{s1}-\hat{\theta}_{s2})}\,.
\end{equation}
This criterion provides a good measure of the stability of the mean (cf. Sec. \ref{sec:CMB}). The Mahalanobis distance is useful because it automatically normalizes the parameters ($D_M$ is dimensionless) and also because it incorporates the information of the existing correlations between them. We can decide to stop the first MC once the distance \eqref{eq:Maha} is lower than some threshold, i.e. once $D_M\lesssim f\sqrt{N}$, with $N$ being the number of parameters and $f$ a number which we can choose to be $f\sim 0.05-0.1$ for a practical application.

The stability of the covariance matrix can be carried out in a similar way. We can obtain first the eigenvalues of $M_{s1}$ and $M_{s2}$. Let us call them $\lambda_i$ and $\beta_i$, respectively, for $i\in [1,N]$. Then we can compute the sum of the absolute values of their relative differences and then demand this quantity to be lower, again, than some threshold,
\begin{equation}\label{eq:distMConv}
d(M_{s1},M_{s2}) \equiv \sum_{i=1}^{N}\left|\frac{\lambda_i}{\beta_i}-1\right|\lesssim\tilde{f} N\,.
\end{equation}
We can take, again, $\tilde{f}\sim 0.05-0.1$. We have explicitly applied these criteria in one of the analyses of Sec. \ref{sec:CMB}. See therein for details.

\end{document}